\documentclass[11pt,a4paper]{article}

\usepackage{graphicx}
\usepackage{epsfig}

\usepackage{hyperref}
\usepackage{graphicx,epsf}
\usepackage{bm}
\usepackage{amsmath}

\begin{document}

\begin{center}
{\large{\textbf{Observation of explosive collisionless reconnection\\ in 3D nonlinear gyrofluid simulations.}}}\\
\vspace{0.4 cm}
{\normalsize {A. Biancalani${}^{1,2}$, Bruce D. Scott${}^1$}\\}
\vspace{0.2 cm}
\small{${}^1$ Max-Planck-Institut f\"ur Plasmaphysik, Euratom Association, D-85748 Garching, Germany\\
${}^2$ Max-Planck-Institut f\"ur Sonnensystemforschung, Katlenburg-Lindau, Germany\\
webaddress of first author: \url{www.ipp.mpg.de/~biancala}}
\end{center}

\begin{abstract}
The nonlinear dynamics of collisionless reconnecting modes is investigated, in the framework of a three-dimensional gyrofluid model. This is the relevant regime of high-temperature plasmas, where reconnection is made possible by electron inertia and has higher growth rates than resistive reconnection.  The presence of a strong guide field is assumed, in a background slab model wih Dirichlet boundary conditions in the direction of nonuniformity. Values of ion sound gyro-radius and electron collisionless skin depth much smaller than the current layer width are considered. Strong acceleration of growth is found at the onset to nonlinearity, while at all times the energy functional is well conserved. Nonlinear growth rates more than one order of magnitude higher than linear growth rates are observed when entering into the small-$\Delta'$ regime.
\end{abstract}

\section{Introduction}

Magnetic reconnection occurs in space and laboratory plasmas, when the equilibrium  magnetic field lines break due to non ideal effects and reconnect with different magnetic topology~\cite{Furth63}. Here, we consider the regime of high-temperature plasmas in the presence of a strong guide magnetic field, where reconnection is made possible by electron inertia and has higher growth rates than resistive reconnection. Such a regime is named collisionless reconnection~\cite{Edwards86,Porcelli91}.

Two dimensional studies of collisionless reconnection with gyrofluid models were performed in Ref.~\cite{Scott04} with code REC2, in a periodic configuration with flat temperature and density equilibrium profiles, and more recently in Ref.~\cite{BiancalaniEPS10} and Ref.~\cite{Grasso10}.
Development of very narrow current layers was also found in the nonlinear phase of collisionless reconnection~\cite{Grasso10,Ottaviani93,Grasso01,Borgogno05}. This suggests the importance of affording high spatial resolution to investigate small scale dynamics, which is crucial in the nonlinear phase of the magnetic island growth. Three dimensional studies with two-fluid model in periodic configuration were also performed in a large-$\Delta'$ regime~\cite{Borgogno05} (where $\Delta'$ is a parameter measuring the ratio of mode wavelength and current layer width). Large-$\Delta'$ regimes, corresponding to large wavelengths, were often adopted in past investiagations for their apparent property of having higher growth rates, and therefore of explaining better fast reconnection observed in nature. Nevertheless, a comprehensive theoretical model capable of reproducing such observed high growth rates does not exist at present day.

Nonlinear growth acceleration was analytically predicted for the m=1 mode in tokamaks, corresponding to large $\Delta'$ regimes in slab model~\cite{Aydemir92}. On the contrary, a nonlinear subexponential growth was predicted analytically for the collisionless tearing mode, corrisponding to small $\Delta'$ regimes in slab model~\cite{DrakePRL77}. Numerically however, a nonlinear growth acceleration was found in fluid simulations of collisionless reconnection~\cite{Grasso99} even for small $\Delta'$ regimes~\cite{Bhattacharjee05}. The nonlinear growth rate was shown to increase for smaller values of the ion sound gyro-radius $\rho_s= c_s / \omega_{ci}$ - with $c_s = \sqrt{T_e/m_i}$ the sound speed and $\omega_{ci}$ the ion cyclotron frequency - and of the electron collisionless skin depth $d_e = c / \omega_{pe}$ - with $c$ the light speed and $\omega_{pe}$ the electron plasma frequency. The value of the  nonlinear growth rate was found up to twice the value of the linear growth rate.

In this paper, we investigate the small-$\Delta'$ regime in the unexplored limit of very small $\rho_s / L_\perp$ and $d_e / L_\perp$, where $L_\perp$ is the current width: the equilibrium characteristic  scale length in the direction of nonuniformity. No gradients of the equilibrium pressure and temperature are set, therefore turbulence is not driven from the beginning of the simulations, but it is allowed to develop selfconsistently with the island evolution during subsequent phases of the simulations.

\section{Model}

Numerical experiments are carried out with the GEM-code (Gyrofluid ElectroMagnetic~\cite{Scott05pop}). This code, originally written for drift wave turbulence in tokamaks, adopts an electromagnetic full-finite-Larmor-radius gyrofluid model of equations for electrons and ions.
Temperature fluctuations are included in the model. The GEM-code has been benchmarked with results of Ref.~\cite{Scott04} and with analytical theory in large $\Delta'$ regime given by Ref.~\cite{Porcelli91}.
The gyrofluid model is obtained by taking the first moments of the gyrokinetic equations~\cite{Dorland93,Scott2000}, with the advantage of incorporating finite ion gyroradius effects at arbitrary order. 
The zeroth and first moments of the gyrokinetic equation for ion and electron species (labeled here by $z = i, e$) are:
\begin{eqnarray}
 \frac{d \tilde{n}_z }{dt} +   {\bf{w}}_E \cdot {\bf{\nabla}}  \tilde{T}_{z\perp} = - B \nabla_\| \frac{\tilde{u}_{z\|}}{B}   \\
 \beta_e \frac{\partial\tilde{A}_\|}{\partial t} + \mu_z \frac{d \tilde{u}_{z \|}}{d t} + \mu_z  {\bf{w}}_E \cdot {\bf{\nabla}} \tilde{q}_{z\perp} =
-\nabla_\| (\tilde{\phi}_G + \tau_z  \tilde{p}_{z\|})  
\end{eqnarray}
where the tilde symbol denotes perturbed gyrocenter densities $\tilde{n}_z$ and velocities $\tilde{u}_z$, scalar potential $\tilde\phi$ and vector potential ${\tilde{\bf{A}}}$, and temperatures $\tilde{T}_z$ (cf. Eqs: 34-39, 83, 86, 99-105 of Ref.~\cite{Scott05pop}). The advective time derivative is given by $d/dt = \partial / \partial t + {\bf{u}}_E\cdot{\bf{\nabla}}$. The time is normalized in terms of $L_\perp / c_s$, with the perpendicular scale length normalized to $\rho_s$, and the parallel gradient $\nabla_\|$ normalized in terms of $L_\perp$. Therefore, the size of the contravariant magnetic unit vector component $b^s$ is comparable to $\hat\epsilon^{-1/2}$, where  $\hat\epsilon = (q R / L_\perp)^2$ and $q R$ and $L_\perp$ are the scale lengths along and perpendicular to the equilibrium magnetic field.
The parameters $\tau_z$, $\beta_e$ and $\mu_z$  are defined by: $\tau_z=T_z/ (Z T_e)$, $ \beta_e  = 4\pi p_e /B^2$ and $ \mu_z =  (m_z/(Z m_i))$. The parallel gradient is calculated in the direction parallel to the total (equilibrium plus perturbed) magnetic field, and the gyro-averaged potentials are defined as $\tilde\phi_G = \Gamma_0^{1/2} (\tilde\phi)$ and $\tilde\Omega_G = T_\perp \partial \tilde\phi_G / \partial T_\perp$.  The fields are advected with the the $E\times B$ drift velocity ${\bf{u}}_E =  c (B \times \nabla \tilde\phi_G ) / B^2$ and with ${\bf{w}}_E =  c (B \times \nabla \tilde\Omega_G ) / B^2$
(where for the electrons we use the bare potential $\tilde\phi_G = \tilde\phi$ and $\tilde\Omega_G = 0$). The FLR nonlinearities are represented both by the difference between $\phi_G$ and $\phi$, and also by ${\bf{w}}_E$. The electromagnetic fields are linked to the sources by the polarization and induction equations:
\begin{eqnarray}
\Gamma_0^{1/2} \tilde{n}_i + \frac{\partial \Gamma_0^{1/2}}{\partial \log T_\perp} \tilde{T}_i   +   \frac{\Gamma_0 - 1}{\tau_i}\tilde{\phi} & = & \tilde{n}_e \\
-\nabla_\perp^2 \tilde{A}_\parallel = \tilde{J}_\parallel &= & \tilde{u}_\parallel - \tilde{v}_\parallel
\end{eqnarray}
where the Pad\'e approximant forms are adopted~\cite{Dorland93}: $\Gamma_0 = (1 - \rho_i^2 \nabla_\perp^2)^{-1}$, $\Gamma_0^{1/2} = (1 - \rho_i^2 \nabla_\perp^2 /2)^{-1}$, and $\rho_i = v_{th,i}/\omega_{ci}$ is the ion gyro-radius ($v_{th,i}$ here is the ion thermal velocity). Note that in the fluid limit, the gyrofluid equations are equivalent to a two-fluid model. As basic assumptions of this model, we neglect fast magneto-sonic waves, and we consider only frequencies smaller than the ion cyclotron frequency (therefore no whistlers are present in this model).
The characteristic scale lengths of this model are the equilibrium current width $L_\perp$, the ion gyro-radius $\rho_i$, the ion sound gyro-radius $\rho_s$, and the collisionless skin depth $d_e$. The characteristic velocities are the Alfv\'en velocity $ v_A = \sqrt{B^2/ 4\pi n_i m_i}$ and the sound speed $ c_s = \sqrt{T_e/m_i}$, where $n_i$ is the ion particle density and $m_i$ the ion mass.

\begin{figure}[t!]
\begin{center}
\includegraphics[width=0.4\textwidth]{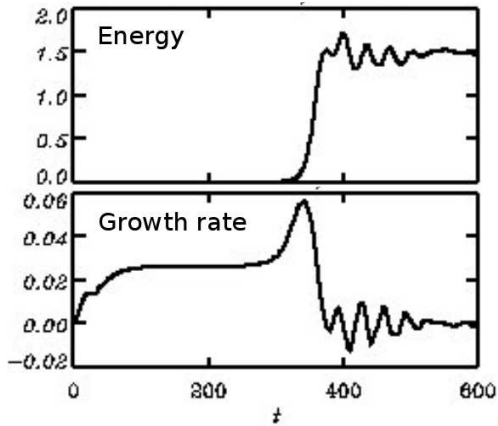}
\caption{ Time evolution of the magnetic island energy (upper figure) and growth rate (lower figure). The time is normalized to $c_s/L_\perp$ and the energy is in arbitrary units. The reconnection growth rate is calculated as the time derivative of the logarithm of the magnetic island energy. Here the plasma parameters are $\beta_e = 5.45\cdot 10^{-5}$, $\rho_i = \rho_s = 0.45 \, d_e$. The nonlinear acceleration is found in this example in the phase $ t \simeq 300$-$350$, and the peak of the nonlinear growth rate is about double of the linear one.}
\label{fig:ETGT_island_seed}
\end{center}
\end{figure}

We consider the plasma regime with $\Delta' \rho_s = 0.75$, and values of $d_e /\rho_s$ ranging from 0.41 to 4.1. We obtain a very thin reconnection layer with respect of the equilibrium current layer ($d_e / L_\perp = 0.04$). This requires a high spatial resolution: we use a grid with $N_x$ up to 2048 points in the perpendicular direction and to 512 in the direction of the sheared field $y$. This also results in a low reconnecting growth rate, requiring very long runs. The parellization of the GEM-code into 64 processors solves the problem of the excessive time length of the runs, bringing the characteristic run time to a few days for the longest runs. A deuterium plasma is considered ($\mu_e = 0.00027$) in our simulations. The value of the parallel to perpendicular scale ratio has been chosen as $\hat\epsilon = 18350$. The ion to electron temperature ratio $\tau_i$ is considered unitary: $\tau_i = 1$.

The evolution of a magnetic island is studied in a Harris-pinch equilibrium, where the equilibrium magnetic field is a hyperbolic tangent of the nonuniformity coordinate $x$. We initialize our simulations with a periodic perturbation with wavenumber $k_y = 2\pi/L_y$, where $L_y$ is the box size in the direction of the shear magnetic field. We add to this initial condition a white noise with amplitude ten times smaller than the magnetic island perturbation. The white noise has components in all three directions, and allows the system - whose equilibrium is two-dimensional - to develop structures in the third direction too. Periodic boundary conditions are imposed in $y$ and in the direction of the equilibrium magnetic field $s$, and Dirichlet boundary conditions in $x$. The choice of the boundary conditions in $x$ differs from those of Ref.~\cite{Bhattacharjee05}, where periodic buondary conditions are chosen also in the $x$ direction. The energy of the nonzonal components of the fluctuating fields is measured and the growth rate is calculated as the time derivative of the energy (see Fig.~\ref{fig:ETGT_island_seed}): $G_T = (dE_T/dt)/(2E_T)$. Such boundary conditions allows free energy to enter from the borders of the simulation box by means of a nonzero gradient of the scalar potential in the $x$ direction, corresponding to ExB drifts in the $y$ directions. The consistency of our choice of boundary conditions has been checked by doubling the size of the box in the perpendicular direction $x$, and finding that the dynamics under investigation is not changed.

\section{Results}

\begin{figure}[t!]
\begin{center}
\includegraphics[width=0.5\textwidth]{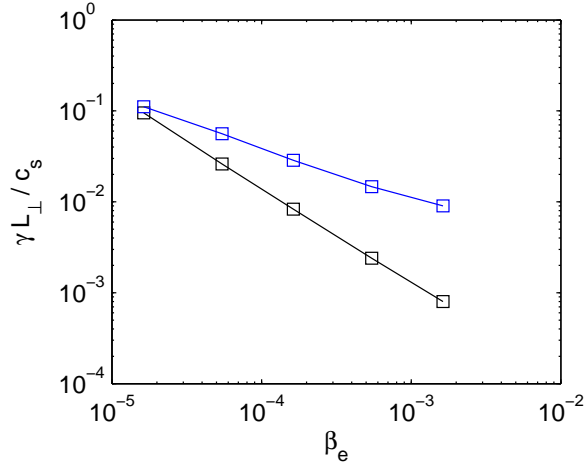}
\caption{ Nonlinear growth rate (upper points, blue) and linear growth rate (lower points, black) normalized with the sound time $L_\perp/c_s$, versus the electron beta $\beta_e$. In these different cases, $d_e$ is the same and the other characteristic lengths are rescaled keeping constant their ratio with $\rho_s$ (therefore the sound time does not depend on $\beta_e$). Explosive reconnection is observed for values of beta of the order of $\beta_e \sim 10^{-3}$ (corresponding to small-$\Delta'$ regime), where the nonlinear growth rate is more than one order of magnitude higher than the linear growth rate.}
\label{fig:gamma_nonlin_vs_beta}
\end{center}
\end{figure}

The initial perturbations are let evolve in time, and different phases in the evolution of the magnetic island are found, defined according to the functional behavior of the growth rate. During an initial transient phase, fluctuations with all wave numbers interact with the magnetic island seed. After the transient phase, a clear linear phase is found, where the magnetic island grows exponentially in time and the growth rate is constant (t $\simeq$ 100 to t $\simeq$ 300 in Fig.~\ref{fig:ETGT_island_seed}). At the onset to the nonlinear phase, the magnetic island amplitude is found to grow superexponentially (t $\simeq$ 300 to t $\simeq$ 350 in Fig.~\ref{fig:ETGT_island_seed}). During this acceleration phase, the growth rate increases, reaches a peak value, whose dependence on $\beta_e$ is shown in Fig.~\ref{fig:gamma_nonlin_vs_beta}, then decreases again. After the nonlinear acceleration phase, the island reaches a saturated amplitude and nonlinearly oscillates around the saturated amplitude. This corresponds to a nonlinear oscillation of the growth rate around the null value. During the saturation phase, the energy flows to secondary instabilities which are formed around the island separatrix.

Simulations are repeated with different plasma parameters, and the value of the peak of the nonlinear growth rate is measured in the different cases. The nonlinear acceleration is found to become more prominent for higher values of $L_\perp / d_e$. In particular, we repeat the simulation as in the case of Fig.~\ref{fig:ETGT_island_seed}, by varying $\beta_e$ and mantaining the same ratio of $L_x/\rho_s$, $L_y/\rho_s$ and $L_\perp/\rho_s$.
In other words, for higher $\beta_e$, we are choosing bigger box sizes with bigger equilibrium current sheet's widths, whereas we keep the same value of $d_e$. The separation in these different phases is found clearly in all these cases with different beta: we always find an initial transient phase, a linear phase, and a nonlinear acceleration followed by a saturation. The value of the peak of the linear and nonlinear growth rates normalized with the sound time $L_\perp / c_s$ for these different cases is shown in Fig.~\ref{fig:gamma_nonlin_vs_beta}. 

The linear growth rate is shown to decrease with a constant slope in the logarithmic graph for increasing $\beta_e$, corresponding to a well defined power law. Increasing the value of $\beta_e$ by one order of magnitude, the linear growth rate decreases also by roughly one order of magnitude. This is consistent with analytical predictions, describing the reconnection process as occurring in a thin ``inertial'' layer with relative width comparable with $d_e/L_\perp$. The smaller the value of $d_e/L_\perp$, the less effective is reconnection. The peak value of the nonlinear growth rate is a measure of how fast the reconnection process occurs during the burst phase. In Fig.~\ref{fig:gamma_nonlin_vs_beta} we show that, contrary to the linear growth rate, the nonlinear growth rate dependence on $\beta_e$ becomes weaker and weaker for higher $\beta_e$. This means that the nonlinear acceleration becomes more relevant the higher is the value of beta. This also means that reconnection process has a tendency to become independent on the microscopic scale size. Increasing the value of $\beta_e$ by two orders of magnitude, the nonlinear growth rate decreases by only one order of magnitude. For computational constraints, the maximum value of $\beta_e$ that we can investigate at present time is $\beta_e = 1.6 \cdot 10^{-3}$. For this value of $\beta_e$, the nonlinear growth rate is $\gamma_{NL} L_\perp / c_s = 0.009$, to be compared with the linear growth rate $\gamma_L L_\perp / c_s = 0.0008$. We call such a regime explosive reconnection.

The evolution of the fields during the nonlinear acceleration phase has been investigated, and some markers have been recognized as characteristic of this particular phase. The analysis of their time evolution shows that they evolve with the same characteristic time scales of the magnetic island, with no higher frequency component. At the O-point, secondary vorticity structures are found by taking a cut along the line $y=0$, $s=0$. The vorticity structure during the whole linear evolution has definited sign on each side of the plane $x=0$, consistent with a uniform inflating of the magnetic island. On the other hand, we find an inversion of sign inside the magnetic island starting with the end of the linear phase. In the case of $\beta_e = 5.45\cdot  10^{-5}$, $\rho_i = \rho_s = 0.45 d_e$, the linear vorticity structure is doubled and reproduced inside the separatrix during the nonlinear acceleration phase, while the growth rate passes from the linear value to double of the linear value. At the X-point position, the formation of a Sweet-Parker like current sheet is found, corresponging to a structure with two Y-points. A successive recovering of the X-point structure is found at the beginning of the saturation phase.
We find that current layers form during the linear phase along the whole separatrix, and their amplitude and steepness increase strongly during the nonlinear acceleration phase. The end of the acceleration phase is characterized by a saturation of the current spike amplitude, and by the formation of new current structures inside the magnetic island.  Acceleration in the magnetic island growth are also found to be linked to the presence of turbulence around the X-point, having the possible effect of multiplying the number of X-points and consequently the rate of reconnection (similarly to results shown in Ref.~\cite{Loureiro09MNRAS}).

The ion finite Larmor radius (FLR) effects are studied by varying the ion to electron temperature ratio $\tau_i$. We find that in the linear phase the FLR effects yields a slightly faster growth rate, consistently with the analytical predictions~\cite{Porcelli91}. The same is valid also in the nonlinear phase. Nevertheless, no qualitative difference is found in the case of $\tau_i=0$ and $\tau_i=1$. The reason is that ion moments don't enter the physics for $J'$-driven processes at low-beta regimes~\cite{Scott04}, namely when $k_\| v_A \gg k_\| c_s$, and for small $\Delta' \rho_s$. On the other hand, FLR effects were found to be more relevant for 2D models in large $\Delta'$ regimes~\cite{Grasso01}.

\section{Energy consistency}

The check of energy consistency is of crucial importance in reconnection simulations. In fact, reconnection processes can occur in numerical experiments also because of artificial - nonphysical - resistivity, such as numerical dissipation. In general, an artificial resistivity causes that the terms of Ohm's law calculated separately don't balance exactly in the subsequent time steps. Therefore, a numerical dissipation can act as a virtual resistivity and the result is a fictitious reconnection, whose growth rate can be in some cases higher than the physical one, depending on the chosen plasma parameters.  In nonlinear simulations,  smaller and smaller scales are created due to nonlinear coupling of macroscopic modes, and it's important to avoid the accumulation of energy at the scales of the numerical grid. To this aim, hyperviscosity is added to our model equations, to filter small scales physics out, just above the grid size.

The contribution to the reconnection growth rate of subgrid dissipation, given by hyperviscosity and numerical dissipation, is calculated with a specific diagnostic in the GEM-code. 
This error growth rate $G_E$ is defined by:
\begin{equation}\label{eq:error-growth-rate}
G_{E} = G_T - G_{So} + G_{Si}
\end{equation}
Here $G_T$ is the reconnection growth rate, calculated as the time derivative of the total energy of the system: $G_T = (dE_T/dt)/(2E_T)$. The total energy is the volume integral of the sum of the energy densities $U_E$, $U_t$, $U_v$ and $U_m$, being respectively the ExB energy, the thermal state variable energy, the flux variable energy and the magnetic energy~\cite{Scott05pop}. $G_{So}$ is the sum of the source growth rates: $G_J = \mu_e v_\| v_E \nabla J_0$, $G_a = - \phi \nabla_\| J$ and $G_r = p_e \nabla_\| J$. 
$G_{Si}$ is the sum of the absolute value of the damping rates, namely Landau damping of ion and electrons (where the electron Landau damping is the relevant one for our regime). The contribute of the numerical error $G_E$ is checked to be always negative during the whole run. In this case, the contribute of the subgrid dissipation to the reconnection rate is null.

\section{Conclusions}

In summary, we have presented the results of simulations on collisionless reconnection in a 3D Harris-pinch equilibrium, with flat density and temperature initial profiles. The gyrofluid code GEM has been used, and single helicity reconnecting modes have been considered in a small $\Delta'$ plasma regime. The limit of small $\rho_s/L_\perp$, $d_e/L_\perp$ has been investigated. The acceleration phase during the nonlinear growth has been analyzed and the nonlinear growth rate peak scaling has been provided, with respect to the equilibrium plasma parameters. Numerical evidence of explosive reconnection has been found for $\beta_e \sim 10^{-3}$. The underlying mechanism of our observation is thought to be linked with the presence of the high velocity fields. In fact, a collisionless regime allows the formation of strong velocity fields during the nonlinear acceleration phase. These velocity fields are retained to be responsible for the splitting of the X-point into two Y-points, and as a consequence, the explosive regime is entered. 
The presence of an acceleration in the growth of the magnetic island, could be important in explaining the fast growth of magnetic islands as seen in nature.

Simulations have been repeated with double or half amplitude of the initial magnetic island seed, and no difference in the value of the nonlinear growth rate has been found. Some markers of the acceleration phase have been pointed out. At the O-point, the formation of smaller vorticity structures inside the magnetic island has been found. At the X-point position, the X-point has been shown to split into two Y-points during the acceleration phase, and then to become again an X-point during the saturation phase. Along the whole separatrix, current spikes at scales narrower than the collisionless skin depths have been found. The importance of these effects in causing the reconnection acceleration is under investigation, and a detailed analysis will be published in a dedicated paper~\cite{BiancalaniPoP2012}.

It is not quite understood whether the three-dimensionality of the configuration is a key ingredient for nonlinear acceleration. In fact, 2D reconnection is known to be governed by conservation laws: being $\phi$ and $A$ functions of the perpendicular variables x and y only, then the current $J$ and the vorticity $\Omega$ are directed in the direction parallel to the guide field, and there is no interaction between neighbour helicities. On the other hand, in 3D simulations neighbour helicities can interact  - for instance J can interact with $\nabla p$ and $\nabla \phi$ - and the nonlinear phase is predicted to have a qualitatively different evolution.

\section*{Acknowledgments}
This work was carried out in a collaboration with the Max-Planck institute for solar system research at Katlenburg-Lindau, Germany. This Letter was written in Metz (France). The authors thank J. Buechner and E. Marsch for interesting discussions. Enlightening discussions with F. Pegoraro, F. Califano, D. Del Sarto, T. Ribeiro and F. Zonca are also gratefully acknowledged.

\end{document}